\documentclass{article}

\PassOptionsToPackage{numbers, compress}{natbib}

\usepackage[final]{neurips_2024}

\usepackage[utf8]{inputenc} 
\usepackage[T1]{fontenc}    
\usepackage{hyperref}       
\usepackage{url}            
\usepackage{booktabs}       
\usepackage{amsfonts}       
\usepackage{nicefrac}       
\usepackage{microtype}      
\usepackage{xcolor}         
\usepackage{caption}

\usepackage{amsmath}
\usepackage{bm}
\usepackage{graphicx}

\title{Accelerating Quantum Emitter Characterization with Latent Neural Ordinary Differential Equations}

\author{%
Andrew H.~Proppe$^{1*}$ \quad Kin Long Kelvin Lee$^{2}$ \quad Weiwei Sun$^{1}$ \quad Chantalle J. Krajewska$^{1}$ \\
\textbf{Oliver Tye}$^{1}$ \quad \textbf{Moungi G. Bawendi}$^{1}$\\
$^1$Department of Chemistry, Massachusetts Institute of Technology, Cambridge, MA 02139 \\
$^2$Accelerated Computing Systems and Graphics, Intel Corporation, Hillsboro, OR 97124\\
\texttt{\{wwsun17, zcqsckr, olijt, mgb\}@mit.edu}\\
\texttt{aproppe@uottawa.ca}
\texttt{kin.long.kelvin.lee@intel.com}
}

\begin{document}

\maketitle

\begin{abstract}
  {Deep neural network models can be used to learn complex dynamics from data and reconstruct sparse or noisy signals, thereby accelerating and augmenting experimental measurements. Evaluating the quantum optical properties of solid-state single-photon emitters is a time-consuming task that typically requires interferometric photon correlation experiments, such as Photon correlation Fourier spectroscopy (PCFS) which measures time-resolved single emitter lineshapes. Here, we demonstrate a latent neural ordinary differential equation model that can forecast a complete and noise-free PCFS experiment from a small subset of noisy correlation functions. By encoding measured photon correlations into an initial value problem, the NODE can be propagated to an arbitrary number of interferometer delay times. We demonstrate this with 10 noisy photon correlation functions that are used to extrapolate an entire de-noised interferograms of up to 200 stage positions, enabling up to a 20-fold speedup in experimental acquisition time from $\sim$3 hours to 10 minutes. Our work presents a new approach to greatly accelerate the experimental characterization of novel quantum emitter materials using deep learning.}
\end{abstract}

\section{Introduction}

Rapid advances in machine learning have enabled the development of new techniques for accelerating scientific measurement. This is especially useful in the domain of quantum mechanics, where the signal is often sparse and dictated by complex underlying physics \cite{gray_machine-learning-assisted_2018, ahmed_quantum_2021, kudyshev_rapid_2020}. A notable example is the characterization of solid-state quantum emitters: such single-photon sources are central to the realization of quantum photonic technologies in areas such as quantum communication \cite{bouchard_achieving_2021}, sensing \cite{lawrie_quantum_2019}, and computation \cite{madsen_quantum_2022}.

Characterizing the quantum optical properties of individual emitters typically requires time-consuming interferometric photon correlation techniques. Photon correlation Fourier spectroscopy (PCFS) measures time-resolved lineshapes of single emitters, providing spectral- and temporal-resolution that is not measured in more traditional Hong-Ou-Mandel interferometry \cite{brokmann_photon-correlation_2006, proppe_PRL}. PCFS requires the measurement of photon correlation functions, $g^{(2)}(\tau, t)$, at several (often more than 100) interferometer stage positions, each with acquisition times of up to several minutes to attain sufficient signal-to-noise ratios. This renders the characterization of multiple quantum emitters an arduous and potentially impossible process, especially when the material under study can degrade and decompose on the same time scale as the measurements. Accelerating these experiments would facilitate more rapid characterization of novel quantum emitters like colloidal quantum dots, an emergent class of quantum-light emitting materials \cite{utzat_coherent_2019, proppe_highly_2023}. 

Experimental speedups can be achieved by lowering data acquisition times at the expense of poorer signal-to-noise ratios, which may be overcome using signal reconstruction algorithms. We have previously shown that this particular class of $g^{(2)}(\tau, t)$ data are difficult to fit with maximum likelihood algorithms, but a machine learning approach---using a deep ensemble of convolutional autoencoders---could be used to denoise both simulated and experimental data, allowing temporal resolution at previously inaccessible timescales of 10 - 100 ns \cite{proppe_PRL}. However, this methodology still required measurement of the $g^{(2)}(\tau, t)$ at every interferometer position. In this work, we take a new approach to accelerating photon correlation experiments using neural ordinary differential equations (NODEs) \cite{chen_neural_2018}. Our key contributions are:

\begin{enumerate}
    \item Development of a proof-of-concept architecture, which a merges conventional recurrent encoder-decoder with the NODE framework, and successfully encodes and predicts the temporal dynamics of entire $g^{(2)}(\tau,t)$ maps,
    \item An additional Fourier loss term, which based on physical intuition, uses the real part of the frequency domain map to help learn oscillatory behavior in the time domain.
\end{enumerate}

We take advantage of the ability of NODEs to be propagated over arbitrary $t$---here corresponding to interferometer delay times---to both interpolate and extrapolate entire PCFS experiments from only a few measured correlation functions. Our model takes as inputs a small subset of measured photon correlation functions $g^{(2)}(\tau, t_{i})$ and encodes them into a latent space, which is then solved over $t$, before being decoded back into the entire $g^{(2)}(\tau, t$) map.
\begin{figure*}[t]
    \centering
    \includegraphics[width=0.96\textwidth]{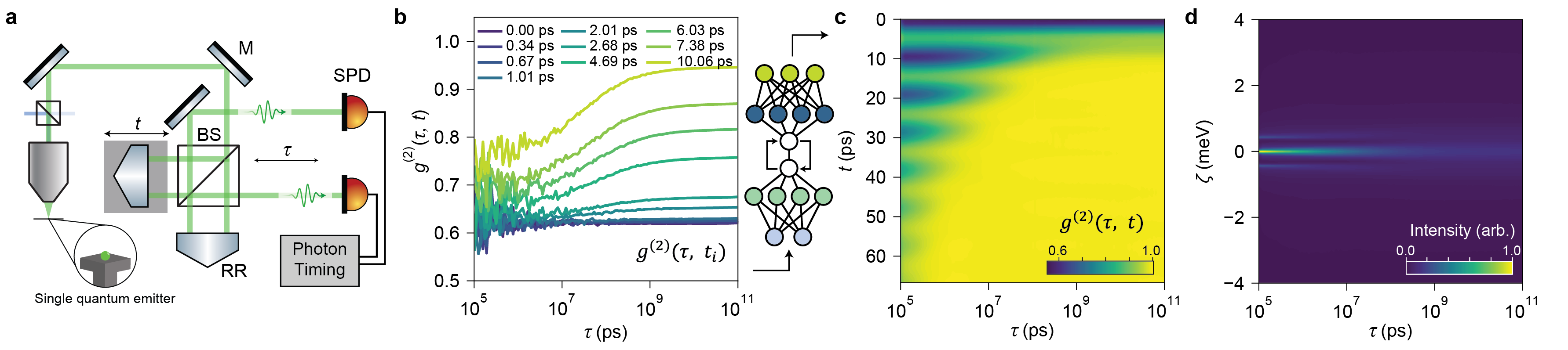}
    \caption{
    ($\boldsymbol{\mathrm{a}}$). Schematic of a PCFS optical setup. BS: beamsplitter, M: mirror, RR: retroreflector, SPD: single-photon detector. ($\boldsymbol{\mathrm{b}}$) Ten $g^{(2)}(\tau, t_{i})$ curves, drawn at values of $t$ shown in the legend, are used as inputs to our model to predict  ($\boldsymbol{\mathrm{c}}$) the full $g^{(2)}(\tau, t)$ map. ($\boldsymbol{\mathrm{d}}$) Fourier transform of the PCFS interferogram (1 - $g^{(2)}(\tau, t)$)  along the $t$ axis, resulting in the spectral correlation $p(\tau, \zeta)$ that shows how the single emitter lineshape evolves over time.
    }  
    \label{fig:fig1}
\end{figure*}
\section{Methods}
\subsection{Photon correlation Fourier spectroscopy}
In PCFS, photons emitted by a single quantum emitter are collected in a microscope objective and sent to an unbalanced Michelson interferometer (Fig.\ref{fig:fig1}a). The outputs of the interferometer are detected by a pair of single-photon detectors. A photon correlation function, $g^{(2)}(\tau, t)$, is measured for each optical delay $t$ (by varying the path length of one interferometer arm). The arrival time between emitted photons arriving at the detectors, $\tau$, provides temporal resolution, whereas scanning the interferometer delay $t$ provides spectral resolution after Fourier transforming along this axis. A detailed explanation of this technique can be found in ref. \cite{brokmann_photon-correlation_2006}. The correlation functions $g^{(2)}(\tau, t)$ are related to the time-dependent spectrum of the emitter by:
\begin{align}
    g^{(2)}(\tau, t) \propto 1 - \mathcal{F}[p(\tau, \zeta)],
    \label{eq:pcfs_g2}
\end{align}
Where $\mathcal{F}$ indicates a Fourier transform, and $p(\tau,\zeta)$ is the autocorrelation of the emission spectrum (with $\zeta$ in units of energy). Several examples of correlation functions at different $t$ are shown in Fig. \ref{fig:fig1}b. Our goal is to use only a fraction of the measured $g^{(2)}(\tau, t_i)$---i.e. along the optical delay $t$ axis---to predict the entire 2D map (Fig.\ref{fig:fig1}c), and from this map obtain the time-resolved lineshapes (Fig.\ref{fig:fig1}d) via eq. \ref{eq:pcfs_g2}. The shot noise is higher at smaller $\tau$ due to fewer coincidence counts occurring in these smaller time bins. In this work, we only consider a $\tau$ range of 0.1 $\mu$s to 100 ms to avoid extremely noisy parts of the data.
\subsection{Neural network models}
The salient feature of our models is a latent NODE layer, which is used in a generative fashion to create solutions at an arbitrary number of timesteps, here corresponding to experimental interferometer delay, $t$. NODEs are continuous-depth models that transform the input by a latent trajectory through an underlying neural network vector field \cite{chen_neural_2018}, 
    $\frac{d\bm{\mathrm{h}}(t)}{dt} = f(\bm{\mathrm{h}}(t), t, \theta)$,
Where $\bm{\mathrm{h}}(t)$ represents the state of the system at time $t$, $f$ is a neural network parameterized by $\theta$, which defines the vector field governing the dynamics, and $\frac{d\bm{\mathrm{h}}(t)}{dt}$ is the time derivative of the state. The NODE vector field is parameterized with multilayer perceptron (MLP) blocks, which propagates a latent space vector obtained from an encoding model which we will describe shortly. For all models, we use the mean-squared error (MSE) between the true and predicted (${g}^{(2)}(\tau, t)$) maps in both the time and frequency domain (discussed later) as the loss function during optimization.

The most performant NODE model in this study uses an encoder-decoder structure: a 3-layer LSTM network (hidden size of 128) encodes the 10 input $g^{(2)}(\tau, t)$ curves. An attention layer combines the LSTM output into a single vector of length $z$, which is used as the initial state $\mathbf{h}(0)$ of the NODE layer. The NODE is propagated for 200 timesteps. This embedding is then decoded by another 3-layer LSTM network (which also receives the hidden state of the encoder) to form the full $g^{(2)}(\tau, t)$ map. We additionally consider NODE models with convolutional encoder and decoder layers, and purely convolutional models with no NODE layer (details in Supplementary Material).

\begin{figure*}[t]
    \centering
    \includegraphics[width=0.96\textwidth]{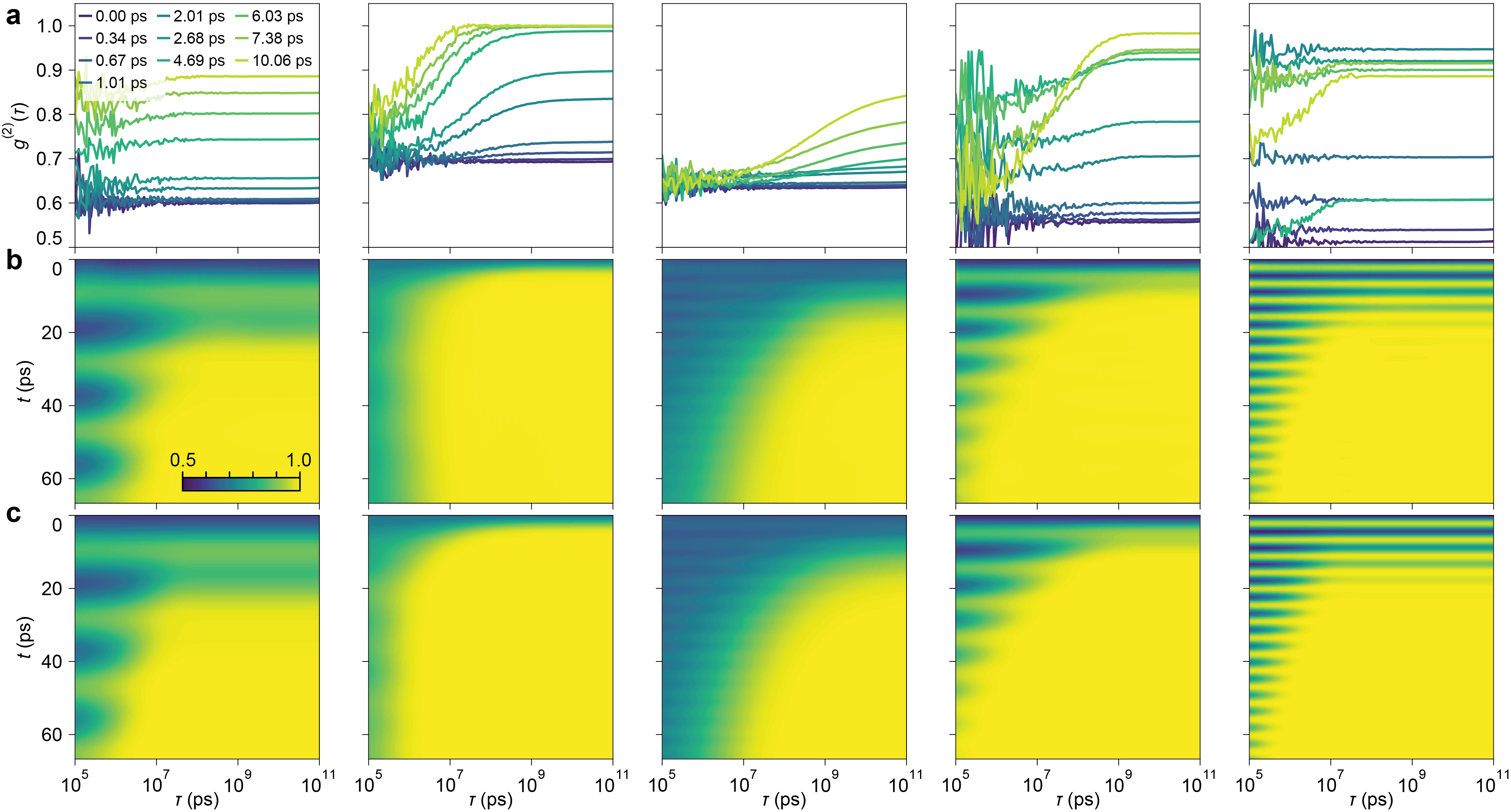}
    \caption{($\boldsymbol{\mathrm{a}}$). Input $g^{(2)}(\tau, t_{i})$ functions at slices of optical delay $t$, ($\boldsymbol{\mathrm{b}}$) $\hat{g}^{(2)}(\tau, t)$ predicted by LSTM-ODE using data in ($\mathbf{a}$), and ($\boldsymbol{\mathrm{c}}$) the true $g^{(2)}(\tau, t$).}
    \label{fig:fig2}
\end{figure*}
\section{Dataset}
We simulate a dataset of 50,000 PCFS $g^{(2)}(\tau, t)$ maps, according to the method developed in ref. \cite{proppe_PRL}. We vary the parameters used to generate the spectra and spectral diffusion mechanisms (see Supplementary Materials for further details). The data are generated by simulating an emission spectrum that diffuses (broadens) over timescale $\tau$, which is auto-correlated to form the spectral correlation $p(\tau,\zeta)$. Fourier transformation of $p(\tau,\zeta)$ yields interferograms that are proportional to $g^{(2)}(\tau, t)$ (see ref. \cite{brokmann_photon-correlation_2006} for details). Noise is added by Poisson sampling to simulate shot noise. This procedure is shown in \ref{fig:figs1}. 10 individual curves at different values of $t$ are drawn from the full 2D map to be used as training inputs. 90\% of the data is used for model training, and 10\% is used for validation. A subsequent set of PCFS maps, generated with a different set of parameters from those used in training and validation, were used for testing. Code for generating data and training models can be found at \url{https://github.com/andrewhproppe/g2-NODE}. The datasets used here can be found at \url{https://zenodo.org/records/13961409}.

\section{Results and Discussion}
Fig. \ref{fig:fig2} shows results for several exemplary LSTM-ODE predictions from the test dataset, demonstrating the model’s performance for a variety of $g^{(2)}(\tau, t)$ shapes resulting from different numbers of emission peaks and spectral splitting (which changes the oscillation patterns), linewidths (which changes the decay rate along $t$) and spectral diffusion mechanisms (which changes the shape of the curves along $\tau$). The training and validation loss curves are shown in Fig. \ref{fig:figs2}. Using only 10 measured $g^{(2)}(\tau, t)$ curves, taken at values of $t$ \emph{that do not exceed 10\,ps}, the model is able to produce reconstructions that strongly resemble the true images across 200 stage positions ranging from $t$ = 0 to 65 ps, with an average MSE $\approx2\times10^{-5}$. We train another LSTM-ODE model with a window of $t$ extended to 120 ps (Fig. \ref{fig:figs3}c), and still observe predictions with similarly low MSE for the predicted $g^{(2)}(\tau, t)$ maps. We also test the performance of models using convolutional encoder-decoder layers, and a 1D ResNet model without a NODE layer (Fig. \ref{fig:figs3}d,e), where we observe inferior performance compared to the LSTM-ODE model (see Supplementary Material for details). The LSTM encoder and decoder layers clearly play an important role to allow the NODE to properly generate the correlation functions along $t$, which capture longer range dependencies than local kernels afforded by convolutions.

We included an additional Fourier loss term in our model training which takes the MSE of the real part of the Fourier transform of the maps along the $t$ axis (i.e. the spectral correlations, $p(\tau, \zeta)$):
\begin{align}
    \mathcal{L}_{\mathcal{F}} = \frac{1}{N}\sum_{j}\left(\text{Re}\{\mathcal{F}(1 - {g}^{(2)}(\tau, t))\} - \text{Re}\{\mathcal{F}({(1 - \hat{g}^{(2)}(\tau, t))})\}\right)^{2},
    \label{eq:fourier_loss}
\end{align}
Fig. \ref{fig:figs4} shows predictions for an LSTM-ODE model trained without $\mathcal{L}_{\mathcal{F}}$. The model is generally able to learn the shape of the correlation functions along the $\tau$ (horizontal) axis, but fails to capture the oscillations along the $t$ (vertical) axis beyond the first few timesteps. This $\mathcal{L}_{\mathcal{F}}$ term is thus essential for capturing the oscillatory patterns along $t$, allowing the model to learn longer range oscillations in the time domain by comparing the true and predicted maps in the frequency domain.

Our LSTM-ODE model is capable of producing entire PCFS interferograms from only a subset of input correlation functions. For the proof-of-concept experiments demonstrated above, the timesteps from the span of $t$ used for these 'observed' data points were selected somewhat arbitrarily, and only span the first 10 ps of a 60 or 120 ps window. Despite such a limited temporal window, the model is still capable of capturing many different dynamics: the rapid initial drop of the $g^{(2)}(\tau, t)$ due to incoherent emission (e.g. into phonon sidebands \cite{besombes_acoustic}), oscillations due to interference between multiple emission peaks, and the (approximately) exponential decay that is proportional to the coherent linewidth.

We posit that the ability of our model to extrapolate these dynamics so accurately from only a 10 ps time window may be partially attributed to the fact that dynamics along $\tau$ are the same regardless of $t$ (for a stable, well-behaved emitter). The model thus has sufficient information to learn a variety of spectral diffusion dynamics along $\tau$, and can focus more on learning how the interference contrast changes as a function of $t$ due to oscillations from multiple peaks and the envelope decay due to finite linewidths.

Despite only being trained on simulated data, we expect these models will also perform well on experimentally measured $g^{(2)}(\tau, t)$ data: we have previously shown that deep ensemble autoencoder models trained only on synthetic $g^{(2)}(\tau, t)$ data still allowed denoising of individual experimental correlation functions \cite{proppe_PRL}.

\section{Conclusions}

In this work, we show that neural differential equations are a natural and intuitive tool for accelerating the characterization of solid-state single-photon emitters. Using a small fraction of input measurements, the LSTM-ODE was capable of accurately predicting the temporal behavior of $g^{(2)}(\tau,t)$ maps. With this, we enable the study of materials that would otherwise be impossible to measure due to poor photostability, with the only requirement now to provide a few measurements at early optical delay stages as input to the model.

\clearpage

\bibliographystyle{unsrt}
\bibliography{refs}     

\appendix

\renewcommand{\thefigure}{S\arabic{figure}}
\setcounter{figure}{0}

\clearpage

\section{Supplementary materials}
\subsection{Dataset generation details}
A simulated PCFS experiment is generated by taking a static, homogeneous emission spectrum, and performing a convolution of this spectrum with a lineshape that diffuses over $\tau$. As in ref. \cite{proppe_PRL}, one type of spectrum consists of 1 - 3 Lorentzian peaks with random amplitudes, linewidths, and relative frequency differences. The second type of spectrum consists of 1 - 2 Lorentzian peaks with an acoustic sideband, parameterized by a spectral density function \cite{berkinsky_acsnano}. We consider two well-known spectral diffusion mechanisms that occur for cryogenically cooled colloidal quantum dots: Wiener diffusion and Poisson diffusion \cite{beyler_prl}. The parameters determining the diffusion evolution along $\tau$ are randomly varied for each mechanism. Furthermore, we augment our datasets by taking random linear combinations of two different PCFS experiments to account for instances where more there may be two simultaneous diffusion mechanisms occurring.

The homogeneous emission spectrum is first auto-correlated to form a static spectral correlation, $\rho_h(\zeta)$, and then undergoes a convolution with the diffusing spectral correlation, $\rho_d(\tau, \zeta)$ to form the overall spectral density $\rho(\tau, \zeta)$. We Fourier transform $\rho(\tau, \zeta)$ along the $\zeta$ axis to obtain the interferogram $I(\tau, \delta)$, where $\delta$ is the interferometer delay time. The $g^{(2)}(\tau, t)$ functions are obtain via eq. \ref{eq:pcfs_g2}. To simulate experimental shot-noise, the $g^{(2)}(\tau, t)$ time bins are scaled by amplitudes proportional to the bin widths (which are roughly logarithmic with respect to $\tau$). The scaled curves are Poisson sampled, with the rate at each bin equal to the amplitude, and then rescaled back to their original values. An example of this procedure is shown in the figure below. 

\begin{figure*}[hbt!]
    \centering
    \includegraphics[width=0.96\textwidth]{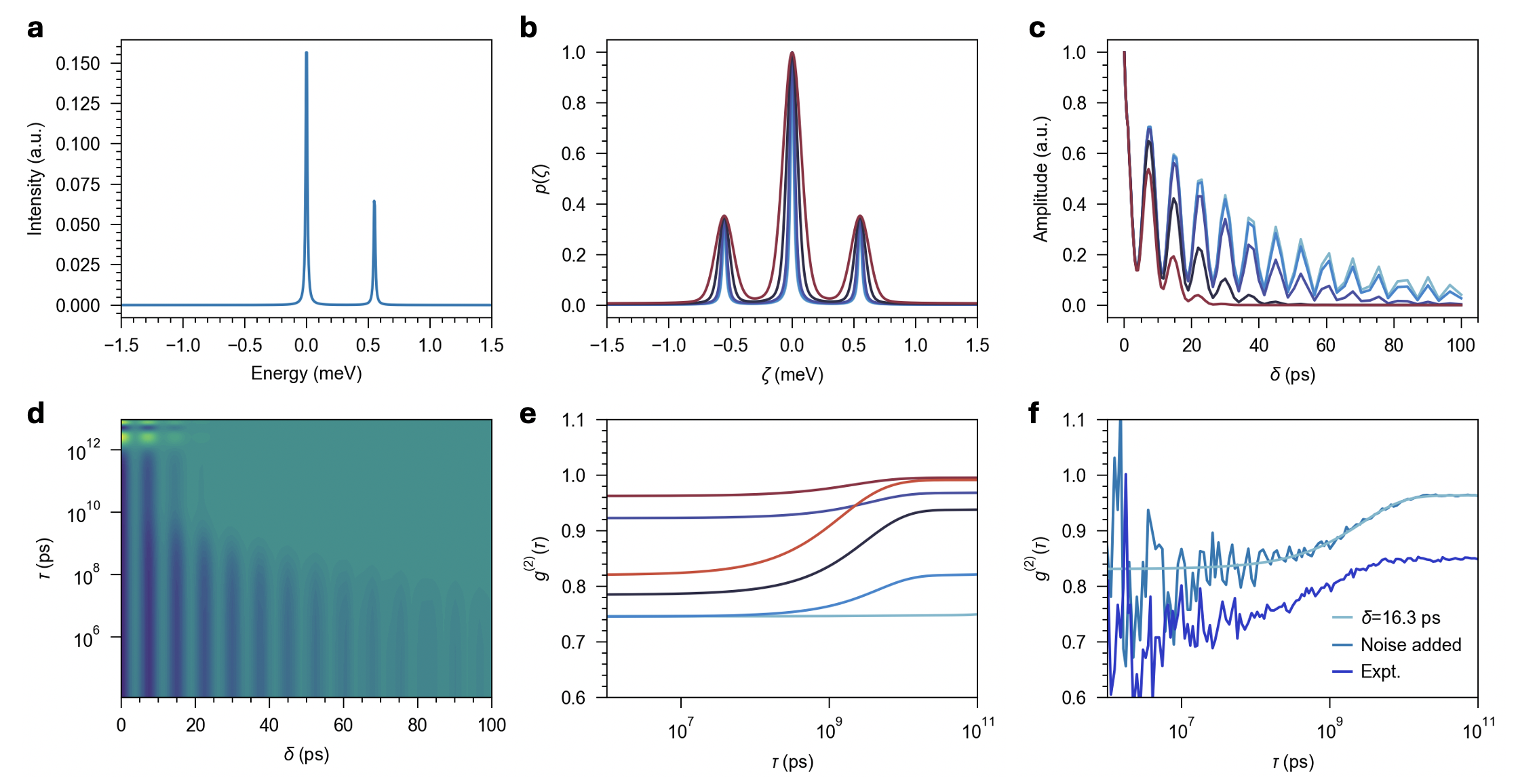}
    \caption{
    (a) Static homogeneous spectrum, (b) spectrum convolved with a diffusing Gaussian that broadens with $\tau$ (using a Wiener diffusion mechanism) to form $\rho(\tau, \zeta)$, (c) Fourier transform of $\rho(\tau, \zeta)$ along $\zeta$ to form the interferogram $I(\tau, \delta)$, (d) $g^{(2)}(\tau, t)$ resulting from the interferogram via eq. \ref{eq:pcfs_g2}, (e) representative individual $g^{(2)}(\tau, t)$ functions along $\tau$ for different values of $\delta$, and (f) an example of a $g^{(2)}(\tau, t)$ curve with and without noise added, compared to a real experimental correlation function.
    }
    \label{fig:figs1}
\end{figure*}

\subsection{Model training loss}

\begin{center}
    \includegraphics[width=0.96\textwidth]{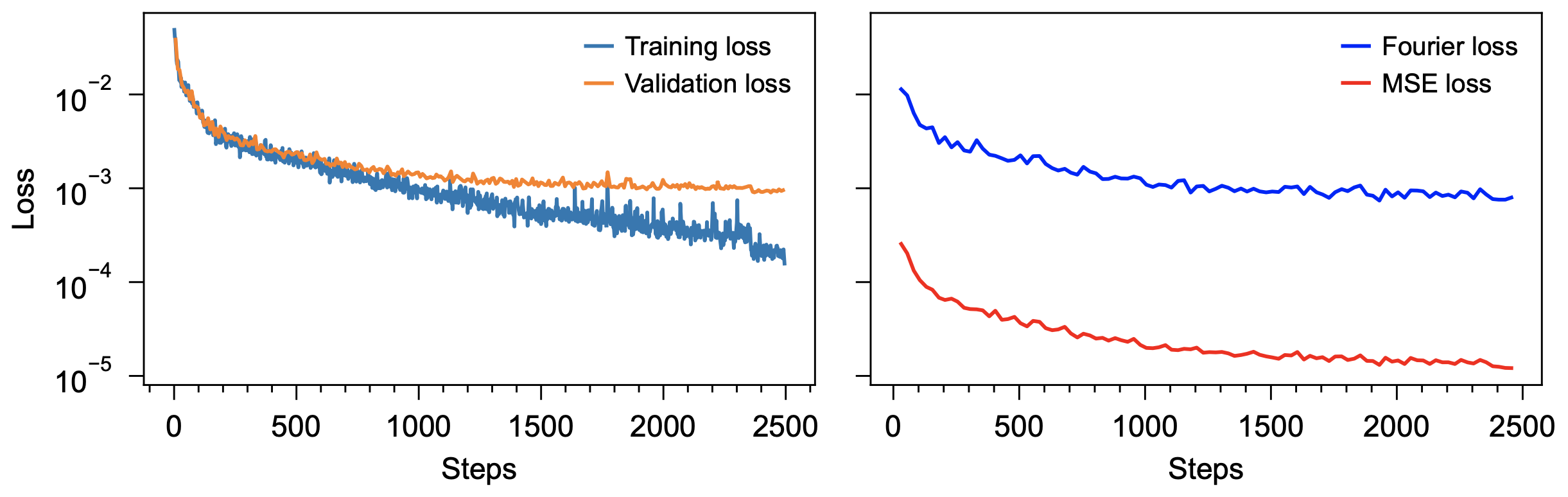}
    \captionof{figure}{
    (left) Training and validation loss curves for the LSTM-ODE model and (right) MSE and Fourier loss components of the validation loss.
    }
    \label{fig:figs2}
\end{center}

\subsection{Alternative neural network models and predictions without Fourier loss term}

We considered two additional models in this study to 1) evaluate the importance of the LSTM encoder/decoder layers surrounding the latent NODE solver, and 2) compare with a model that doesn't use an NODE to generate predictions at unobserved interferometer delay times: 

\setlength{\parskip}{0pt} 
\paragraph{Conv-ODE}
Similar to the above model but with convolutional neural network layers replacing the LSTM encoder and decoder layers. The convolutional encoder is a 2D ResNet using strided convolutions with a kernel size of 3. The decoder is a 1D ResNet of convolutional transpose operations for upsampling the dimensionality along $\tau$ back to the size of the original input.
\paragraph{1D ResNet}
To compare with the above models that generate data at different timesteps by propagating the NODE, we also perform experiments with a 1D residual network that treats the interferometer delays $t$ as convolutional channels, starting with 10 channels and increasing to 200 over 4---8 layers of 1D ResNet blocks with a kernel size of 3. The convolutions are padded and no stride or pooling is used to maintain the $\tau$ dimension.
\setlength{\parskip}{5.5pt}  

Below, we compare predictions from the same set of inputs (panel a) for the LSTM-ODE, Conv-ODE, and 1D ResNet models. The Conv-ODE model completely fails to learn and suffers from reconstruction artefacts at several values of $\tau$. The 1D ResNet appears to be partly successful in predicting the full $g^{(2)}(\tau, t)$ map, but similarly suffers from random missing sections of the reconstruction. 

\begin{figure*}[h!]
    \centering
    \includegraphics[width=0.96\textwidth]{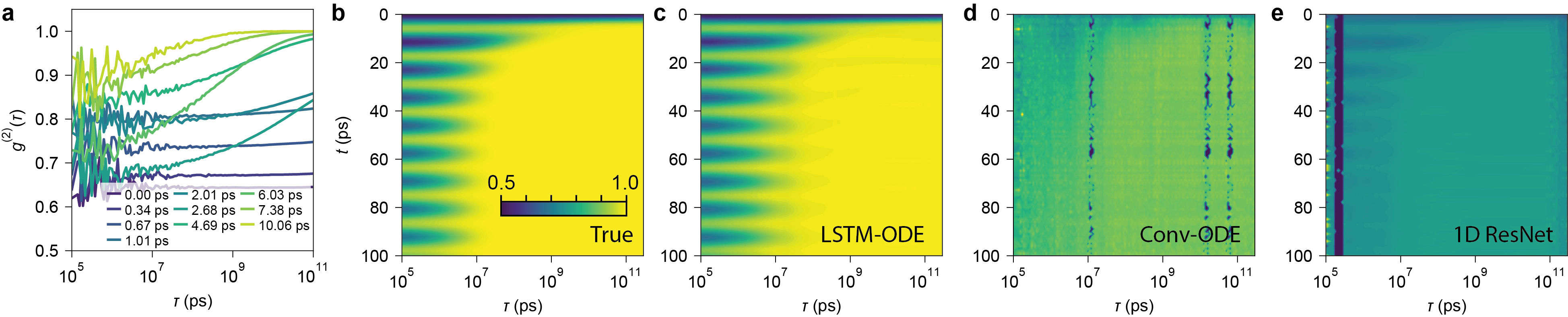}
    \caption{
    ($\boldsymbol{\mathrm{a}}$). Input $g^{(2)}(\tau, t_{i})$ functions at slices of optical delay $t$, ($\boldsymbol{\mathrm{b}}$) True ${g}^{(2)}(\tau, t)$ map, ($\boldsymbol{\mathrm{c}}$) Predicted $\hat{g}^{(2)}(\tau, t)$ from the LSTM-ODE, ($\boldsymbol{\mathrm{d}}$) Conv-ODE and ($\boldsymbol{\mathrm{e}}$) 1D ResNet models.
    }
    \label{fig:figs3}
\end{figure*}

\begin{figure*}[h!]
    \centering
    \includegraphics[width=0.80\textwidth]{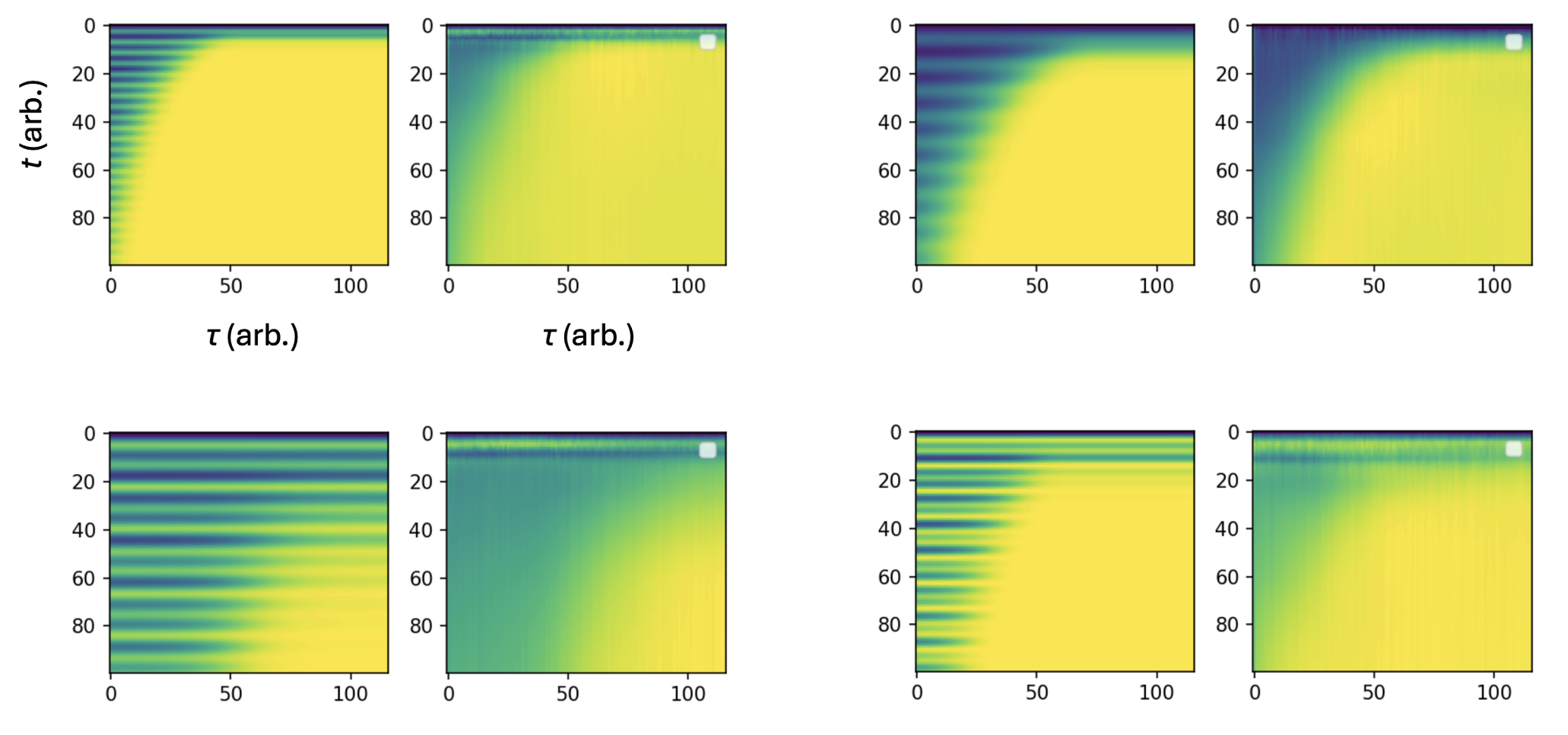}
    \caption{Four examples of true ${g}^{(2)}(\tau, t)$ (left panels) and predicted $\hat{g}^{(2)}(\tau, t)$ (right panels) for LSTM-ODE models trained without the Fourier loss term $\mathcal{L}_{\mathcal{F}}$ (eq. \ref{eq:fourier_loss}). In practice, we found it more memory efficient to only use the Fourier transformed data at the first, middle, and last index of the $\tau$ axis.}
    \label{fig:figs4}
\end{figure*}


\end{document}